\documentclass[aps,prl,twocolumn,superscriptaddress,floatfix,showpacs]{revtex4}
\usepackage{amsmath,fancybox}
\usepackage{graphicx}
\usepackage{epsfig,subfigure}
\begin{document}

\title{Exotic Meson Decay to $\omega\pi^{0}\pi^{-}$}

\author{M.~\surname{Lu}}
\altaffiliation[Present address: ]{Department of Physics, University of Oregon, Eugene, Oregon 97403}
\affiliation{Department of Physics, Rensselaer Polytechnic Institute, Troy, New York 12180}
\author{G.~S.~\surname{Adams}}
\affiliation{Department of Physics, Rensselaer Polytechnic Institute, Troy, New York 12180}
\author{T.~\surname{Adams}}
\altaffiliation[Present address: ]{Department of Physics, Florida State University, Tallahassee, FL 32306}
\affiliation{Department of Physics, University of Notre Dame, Notre Dame, Indiana 46556}
\author{Z.~\surname{Bar-Yam}}
\affiliation{Department of Physics, University of Massachusetts Dartmouth, North Dartmouth, Massachusetts 02747}
\author{J.~M.~\surname{Bishop}}
\affiliation{Department of Physics, University of Notre Dame, Notre Dame, Indiana 46556}
\author{V.~A.~\surname{Bodyagin}}
\altaffiliation{Deceased}
\affiliation{Nuclear Physics Institute, Moscow State University, Moscow, Russian Federation 119899}
\author{D.~S.~\surname{Brown}}
\altaffiliation[Present address: ]{Department of Physics, University of Maryland, College Park, MD 20742}
\affiliation{Department of Physics, Northwestern University, Evanston, Illinois 60208}
\author{N.~M.~\surname{Cason}}
\affiliation{Department of Physics, University of Notre Dame, Notre Dame, Indiana 46556}
\author{S.~U.~\surname{Chung}}
\affiliation{Physics Department, Brookhaven National Laboratory, Upton, New York 11973}
\author{J.~P.~\surname{Cummings}}
\affiliation{Department of Physics, Rensselaer Polytechnic Institute, Troy, New York 12180}
\author{K.~\surname{Danyo}}
\affiliation{Physics Department, Brookhaven National Laboratory, Upton, New York 11973}
\author{A.~I.~\surname{Demianov}}
\affiliation{Nuclear Physics Institute, Moscow State University, Moscow, Russian Federation 119899}
\author{S.~P.~\surname{Denisov}}
\affiliation{Institute for High Energy Physics, Protvino, Russian Federation 142284}
\author{V.~\surname{Dorofeev}}
\affiliation{Institute for High Energy Physics, Protvino, Russian Federation 142284}
\author{J.~P.~\surname{Dowd}}
\affiliation{Department of Physics, University of Massachusetts Dartmouth, North Dartmouth, Massachusetts 02747}
\author{P.~\surname{Eugenio}}
\affiliation{Department of Physics, Florida State University, Tallahassee, FL 32306}
\author{X.~L.~\surname{Fan}}
\affiliation{Department of Physics, Northwestern University, Evanston, Illinois 60208}
\author{A.~M.~\surname{Gribushin}}
\affiliation{Nuclear Physics Institute, Moscow State University, Moscow, Russian Federation 119899}
\author{R.~W.~\surname{Hackenburg}}
\affiliation{Physics Department, Brookhaven National Laboratory, Upton, New York 11973}
\author{M.~\surname{Hayek}}
\altaffiliation[Permanent address: ]{Rafael, Haifa, Israel}
\affiliation{Department of Physics, University of Massachusetts Dartmouth, North Dartmouth, Massachusetts 02747}
\author{J.~\surname{Hu}}
\altaffiliation[Present address: ]{TRIUMF, Vancouver, B.C., V6T 2A3, Canada}
\affiliation{Department of Physics, Rensselaer Polytechnic Institute, Troy, New York 12180}
\author{E.~I.~\surname{Ivanov}}
\affiliation{Department of Physics, Idaho State University, Pocatello, ID 83209}
\author{D.~\surname{Joffe}}
\affiliation{Department of Physics, Northwestern University, Evanston, Illinois 60208}
\author{I.~\surname{Kachaev}}
\affiliation{Institute for High Energy Physics, Protvino, Russian Federation 142284}
\author{W.~\surname{Kern}}
\affiliation{Department of Physics, University of Massachusetts Dartmouth, North Dartmouth, Massachusetts 02747}
\author{E.~\surname{King}}
\affiliation{Department of Physics, University of Massachusetts Dartmouth, North Dartmouth, Massachusetts 02747}
\author{O.~L.~\surname{Kodolova}}
\affiliation{Nuclear Physics Institute, Moscow State University, Moscow, Russian Federation 119899}
\author{V.~L.~\surname{Korotkikh}}
\affiliation{Nuclear Physics Institute, Moscow State University, Moscow, Russian Federation 119899}
\author{M.~A.~\surname{Kostin}}
\affiliation{Nuclear Physics Institute, Moscow State University, Moscow, Russian Federation 119899}
\author{J.~\surname{Kuhn}}
\altaffiliation[Present address: ]{Department of Physics, Carnegie Mellon University, Pittsburgh, Pennsylvania 15213}
\affiliation{Department of Physics, Rensselaer Polytechnic Institute, Troy, New York 12180}
\author{V.~V.~\surname{Lipaev}}
\affiliation{Institute for High Energy Physics, Protvino, Russian Federation 142284}
\author{J.~M.~\surname{LoSecco}}
\affiliation{Department of Physics, University of Notre Dame, Notre Dame, Indiana 46556}
\author{J.~J.~\surname{Manak}}
\affiliation{Department of Physics, University of Notre Dame, Notre Dame, Indiana 46556}
\author{J.~\surname{Napolitano}}
\affiliation{Department of Physics, Rensselaer Polytechnic Institute, Troy, New York 12180}
\author{M.~\surname{Nozar}}
\altaffiliation[Present address: ]{Thomas Jefferson National Accelerator Facility, Newport News, Virginia 23606}
\affiliation{Department of Physics, Rensselaer Polytechnic Institute, Troy, New York 12180}
\author{C.~\surname{Olchanski}}
\altaffiliation[Present address: ]{TRIUMF, Vancouver, B.C., V6T 2A3, Canada}
\affiliation{Physics Department, Brookhaven National Laboratory, Upton, New York 11973}
\author{A.~I.~\surname{Ostrovidov}}
\affiliation{Department of Physics, Florida State University, Tallahassee, FL 32306}
\author{T.~K.~\surname{Pedlar}}
\altaffiliation[Present address: ]{Laboratory for Nuclear Studies, Cornell University, Ithaca, NY 14853}
\affiliation{Department of Physics, Northwestern University, Evanston, Illinois 60208}
\author{A.~V.~\surname{Popov}}
\affiliation{Institute for High Energy Physics, Protvino, Russian Federation 142284}
\author{D.~I.~\surname{Ryabchikov}}
\affiliation{Institute for High Energy Physics, Protvino, Russian Federation 142284}
\author{L.~I.~\surname{Sarycheva}}
\affiliation{Nuclear Physics Institute, Moscow State University, Moscow, Russian Federation 119899}
\author{K.~K.~\surname{Seth}}
\affiliation{Department of Physics, Northwestern University, Evanston, Illinois 60208}
\author{N.~\surname{Shenhav}}
\altaffiliation[Permanent address: ]{Rafael, Haifa, Israel}
\affiliation{Department of Physics, University of Massachusetts Dartmouth, North Dartmouth, Massachusetts 02747}
\author{X.~\surname{Shen}}
\altaffiliation[Permanent address: ]{Institute of High Energy Physics, Bejing, China}
\affiliation{Department of Physics, Northwestern University, Evanston, Illinois 60208}
\affiliation{Thomas Jefferson National Accelerator Facility, Newport News, Virginia 23606}
\author{W.~D.~\surname{Shephard}}
\affiliation{Department of Physics, University of Notre Dame, Notre Dame, Indiana 46556}
\author{N.~B.~\surname{Sinev}}
\affiliation{Nuclear Physics Institute, Moscow State University, Moscow, Russian Federation 119899}
\author{D.~L.~\surname{Stienike}}
\affiliation{Department of Physics, University of Notre Dame, Notre Dame, Indiana 46556}
\author{J.~S.~\surname{Suh}}
\altaffiliation[Present address: ]{Department of Physics, Kyungpook National University, Daegu, Korea}
\affiliation{Physics Department, Brookhaven National Laboratory, Upton, New York 11973}
\author{S.~A.~\surname{Taegar}}
\affiliation{Department of Physics, University of Notre Dame, Notre Dame, Indiana 46556}
\author{A.~\surname{Tomaradze}}
\affiliation{Department of Physics, Northwestern University, Evanston, Illinois 60208}
\author{I.~N.~\surname{Vardanyan}}
\affiliation{Nuclear Physics Institute, Moscow State University, Moscow, Russian Federation 119899}
\author{D.~P.~\surname{Weygand}}
\affiliation{Thomas Jefferson National Accelerator Facility, Newport News, Virginia 23606}
\author{D.~B.~\surname{White}}
\affiliation{Department of Physics, Rensselaer Polytechnic Institute, Troy, New York 12180}
\author{H.~J.~\surname{Willutzki}}
\altaffiliation{Deceased}
\affiliation{Physics Department, Brookhaven National Laboratory, Upton, New York 11973}
\author{M.~\surname{Witkowski}}
\affiliation{Department of Physics, Rensselaer Polytechnic Institute, Troy, New York 12180}
\author{A.~A.~\surname{Yershov}}
\affiliation{Nuclear Physics Institute, Moscow State University, Moscow, Russian Federation 119899}

% =====  End of collaborator's list  =====% Collaboration name% ==================

\collaboration{The E852 collaboration}

% =====  End of collaboration name  =====

\date{\today}

\begin{abstract}
A partial-wave analysis of the mesons from the reaction $\pi^{-}%
p\rightarrow\pi^{+}\pi^{-}\pi^{-}\pi^{0}\pi^{0}p$ has been performed.  The
data show $b_{1}\pi$ decay of the spin-exotic states $\pi_{1}(1600)$ and
\ $\pi_{1}(2000)$.  Three isovector $2^{-+}$ states were seen in the $\omega\rho^{-}$ 
decay channel. 
In addition to the well known $\pi_{2}(1670)$, signals were also observed 
for $\pi_{2}(1880)$ and $\pi_{2}(1970)$. 

\end{abstract}

%\begin{center}
%{\Large \it \textbf{Draft 2.7 --- Please Do Not Copy}}
%\end{center}

\pacs{13.25-k, 13.85.Hd, 14.40.Cs}

\maketitle

Interest in exotic mesons predates the emergence of quantum chromodynamics
(QCD) as the fundamental theory of the strong interaction \cite{bigref}.
\ With the widespread acceptance of QCD one may hope that a study of gluonic
matter will yield insights into the nature of color confinement \cite{Isgur99}.
\ States with manifestly exotic quantum numbers are particularly vital to our
understanding of hadron structure because they cannot have the quark-antiquark 
structure exhibited by most mesons. \ Lattice-gauge calculations 
show that the lightest
of these should be $J^{PC}=1^{-+}$states having a mass around 1.9 GeV$/c^{2}$~\cite{lattice}.

\ Three isovector exotic mesons have recently been discovered. \ An isovector
$1^{-+}$state at 1.4 GeV$/c^{2}$ was reported in $\eta\pi$ decay~\cite{eta,Abele}, 
and another isovector $1^{-+}$ meson, $\pi_{1}(1600)$, was
observed in $\rho\pi$~\cite{rho}, $\eta^{\prime}\pi$~\cite{eta'}, and $f_{1}\pi$~\cite{Kuhn} 
decay.  The latter experiment also revealed a higher state, $\pi_{1}(2000)$~\cite{Kuhn}. \ This
rich spectrum of exotic mesons is somewhat puzzling; lattice \cite{lattice}
and flux-tube model \cite{Isgur,Barnes95} calculations predict only one low-mass  
$\pi_{1}$ meson. \ Glueballs, being pure glue states and hence isoscalar, do not
affect the $\pi_{1}$ spectrum \cite{glueball}.\ Donnachie~\cite{Donnachie} and
Szczepaniak~\cite{Sz} have proposed dynamical origins for $\pi_{1}(1400)$ and/or
$\pi_{1}(1600)$. Four-quark configurations may also contribute to spin-exotic
mesons. \ Further progress in understanding these states, as well as gluonic
mesons with conventional quantum numbers, depends on achieving a better
understanding of their decay properties.

In the flux-tube model the lightest $1^{-+}$ isovector hybrid is predicted to
decay primarily to $b_{1}\pi$ \cite{Isgur}. The $f_{1}\pi$ branch is also
expected to be large and many other decay modes are suppressed. \ This
suppression is consistent with recent calculations showing $1/N_{c}^{2}$
\ behavior for decays to spin-zero mesons in the large-$N_{c}$ limit of QCD~\cite{pageN}.  
Recent refinements in the flux-tube calculations cast some doubt on the previous
estimates of small $\pi_{1}$ branching-widths~\cite{CloseDudek}.  

Few experiments have addressed the $b_{1}\pi$ and $f_{1}\pi$ decay channels.
\ The VES collaboration reported a broad $1^{-+}$\ peak in $b_{1}\pi$
decay~\cite{VES}, and Lee, \textit{et al.}~\cite{Lee} observed significant
$1^{-+}$ strength in $f_{1}\pi$\ decay. \ In neither case
was a definitive resonance interpretation of the $1^{-+}$ waves possible.  Preliminary results from
a later VES analysis show excitation of $\pi_{1}(1600)$~\cite{Dorofeev}.
\ A recent experiment measured
$f_{1}\pi$ decay of $\pi_{1}(1600)$ and $\pi_{1}(2000)$~\cite{Kuhn}.

In this letter we report an analysis of the reaction $\pi^{-}p\rightarrow
\pi^{+}\pi^{-}\pi^{-}\pi^{0}\pi^{0}p.$ \ Partial-wave fits of the mesons from
this reaction show the exotic $\pi_{1}(1600)$ and \ $\pi_{1}(2000)$ states in
$b_{1}\pi$ waves. \ We also observe three isovector $2^{-+}$ resonances, thus 
clarifying the spectroscopy of $\pi_{2}$ mesons~\cite{Barnes}.

The data sample was collected during the 1995 run of experiment E852 at the
Multi-Particle Spectrometer facility at Brookhaven National Laboratory (BNL).
A $\pi^{-}$\ beam, with laboratory momentum 18 GeV$/c$, and a liquid hydrogen
target were used. A description of the experimental apparatus can be
found in Ref.~\cite{eta}.

\begin{figure}
\includegraphics[width=9cm]{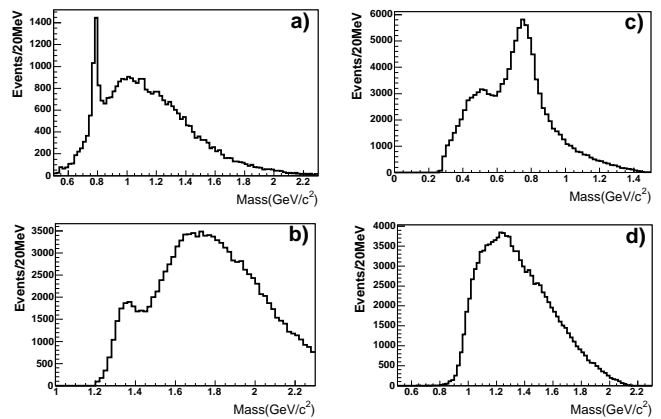}%
\caption{\label{fig:1}Invariant mass of (a) $\pi^{+}\pi^{-}\pi^{0}$ before the $\omega$ mass cut (all
combinations), (b) $\omega\pi^{0}\pi^{-}$, (c) $\pi^{-}\pi^{0}$ using the $\pi^{-}$ and $\pi^{0}$ 
not from the $\omega$,
showing $\rho(770)$, and (d) $\omega\pi^{0}.$}
\end{figure}

Data acquisition was triggered on three forward-going charged tracks, a
charged recoil track, and a signal in a lead-glass electromagnetic calorimeter
(LGD). A total of 165 million triggers of this type were recorded. After
reconstruction, 1.37 million events satisfied the trigger topology and had
four photon-clusters in the LGD. \ Fiducial cuts were then applied on the target and 
detector volumes, and a kinematic
fit \cite{squaw} was performed to select events that were consistent with the
reaction $\pi^{-}p\rightarrow\pi^{+}\pi^{-}\pi^{-}\pi^{0}\pi^{0}p$. \ Events
with confidence level greater than five percent were retained. \ Further
background suppression was achieved by rejecting events for which the measured
proton azimuthal angle differed from that of the missing momentum by more than $20$ 
degrees. \ Finally, events that
were kinematically consistent with $\eta\rightarrow\pi^{+}\pi^{-}\pi^{0}$ detection were
rejected, so as to simplify the partial-wave analysis. \ Those events with
$\pi^{+}\pi^{-}\pi^{0}$ invariant mass near the $\omega(782)$ mass were
selected with a mass cut. \ If more than one mass combination fell in
the cut region (26\% of the sample) a random selection was made between the $\omega(782)$ candidates. \ 
This
process resulted in a final data sample of 145,148 $\omega\pi^{-}\pi^{0}$
events. Mass plots for those data are shown in Figure~\ref{fig:1}.\ 

Figure~\ref{fig:1}(a) shows the $\pi^{0}\pi^{-}\pi^{+}$ mass spectrum for a small sample of the data, 
before $\omega(782)$ selection.  A clear peak is evident at the $\omega(782)$ mass.  
Based on a Monte Carlo simulation of the detector acceptance, we estimate that 
about 21\% of the events that passed the
$\omega$ mass cut did not have an $\omega$ in the final state.  Figures~\ref{fig:1}(b), (c), and (d) 
show mass distributions after $\omega$ selection.  
Evidence for the $\omega\rho^{-}$ (Fig.~\ref{fig:1}(c)) and $b_{1}\pi$ (Fig.~\ref{fig:1}(d)) final 
states is clear.
The $\omega\pi^{-}$ mass distribution (not 
shown) is similar to that for $\omega\pi^{0}$.\ For the
final partial-wave fits a further selection was made on the four-momentum
transfer to the five-pion system $(0.1<-t<1.0$ GeV$^{2}/c^{2})$ and meson 
invariant mass (M $\leq2.2$GeV/$c^{2}$). \ The data follow an $e^{-4.5\left\vert
t\right\vert }$ shape.

A partial wave analysis (PWA) of the present data was made in the isobar model, using 
the maximum likelihood method~\cite{pwa}. Full rank-2 fits were studied with waves in the range
$J\leq4,L\leq3,$ and $m\leq1$, where $J$ is total angular momentum, $L$ is the
decay orbital-angular momentum, and\textit{ }$m$ is the magnitude of the
beam-projection of $J$. \ The mass of the $\pi^{+}\pi^{-}\pi^{-}\pi^{0}\pi
^{0}$ final state was binned in 80 MeV$/c^{2}$ intervals and independent fits
were performed on the data in each bin. The final state was represented as a
sequence of interfering two-body intermediate states. An initial decay of a
parent meson into an intermediate resonance (isobar) and an unpaired meson, or
two isobars, followed by the subsequent decay of the isobars, populates the
final state. The experimental acceptance was determined by means of a Monte
Carlo simulation, which was then incorporated into the PWA normalization for
each partial wave. The same data selection methods that were used for the experimental 
data were also applied to the simulated data.  
Published values were used for the isobar widths~\cite{pdg}.
\ Decays containing more than one charge state for an isobar were constrained
to form a single wave with total isospin equal to one. \ 

Groups of waves were added to the fit in succession, starting with $\omega
\rho^{-}$ and $(b_{1}\pi)^{-}$, and small waves were removed at each
stage. \ Isovector $a_{1}\sigma,a_{2}\sigma,$ and $\rho(1450)\pi$ waves were
also tested and found to be negligible. \ The final set of waves is shown in
Table~\ref{table:1}. \ Isovector $\omega\rho,b_{1}\pi$ and $\rho_{3}(1690)\pi$ waves are
present.%

\begin{table}[tbp] \centering
\caption{Waves in the final fit.  Here positive $\epsilon$ indicates natural parity exchange and $s$ is the total
spin of the initial decay products.  An isotropic background wave was also included.\label{table:1}}
\begin{tabular}
[c]{|l|l|l|l|l|l|l|l|l|l|l|}\cline{1-5}\cline{7-11}
$\bf{decay}$ & $\bf{L}$ & $\mathbf{J}^{pc}$ & $\mathbf{s}$ & $\mathbf{m}^{\epsilon}$ &  &  
$\bf{decay}$ & $\bf{L}$ & $\mathbf{J}^{pc}$ & $\mathbf{s}$ & $\mathbf{m}^{\epsilon}$\\\cline{1-5}
\cline{7-11}%
$\omega\rho $ & $S$ & 1$^{++}$ & 1 & 0$^{+}$ &  & $b_{1}\pi $ & $S$ & 1$^{-+}$ & 1 &
1$^{+}$\\\cline{1-5}\cline{7-11}%
$\omega\rho $ & $S$ & 2$^{++}$ & 2 & 0$^{-}$ &  & $b_{1}\pi $ & $S$ & 1$^{-+}$ & 1 &
1$^{-}$\\\cline{1-5}\cline{7-11}%
$\omega\rho $ & $S$ & 2$^{++}$ & \multicolumn{1}{|c|}{2} & 1$^{+}$ &  & $b_{1}\pi
$ & $S$ & 1$^{-+}$ & 1 & 0$^{-}$\\\cline{1-5}\cline{7-11}%
$\omega\rho $ & $P$ & 0$^{-+}$ & 1 & 0$^{+}$ &  & $b_{1}\pi $ & $P$ & 1$^{++}$ & 1 &
0$^{+}$\\\cline{1-5}\cline{7-11}%
$\omega\rho $ & $P$ & 2$^{-+}$ & 1 & 0$^{+}$ &  & $b_{1}\pi $ & $P$ & 1$^{++}$ & 1 &
1$^{+}$\\\cline{1-5}\cline{7-11}%
$\omega\rho $ & $P$ & 2$^{-+}$ & 1 & 1$^{-}$ &  & $b_{1}\pi $ & $P$ & 2$^{++}$ & 1 &
1$^{+}$\\\cline{1-5}\cline{7-11}%
$\omega\rho $ & $P$ & 2$^{-+}$ & 2 & 0$^{+}$ &  & $b_{1}\pi $ & $P$ & 2$^{++}$ & 1 &
0$^{-}$\\\cline{1-5}\cline{7-11}%
$\omega\rho $ & $P$ & 2$^{-+}$ & 2 & 1$^{+}$ &  & $b_{1}\pi $ & $D$ & 2$^{-+}$ & 1 &
0$^{+}$\\\cline{1-5}\cline{7-11}%
$\omega\rho $ & $D$ & 1$^{++}$ & 2 & 0$^{+}$ &  & $b_{1}\pi $ & $D$ & 2$^{-+}$ & 1 &
1$^{-}$\\\cline{1-5}\cline{7-11}%
$\omega\rho $ & $D$ & 1$^{++}$ & 2 & 1$^{+}$ &  & $b_{1}\pi $ & $D$ & 2$^{-+}$ & 1 &
1$^{+}$\\\cline{1-5}\cline{7-11}%
$\omega\rho $ & $D$ & 3$^{++}$ & 2 & 0$^{+}$ &  & $b_{1}\pi $ & $F$ & 2$^{++}$ & 1 &
1$^{+}$\\\cline{1-5}\cline{7-11}%
$\omega\rho $ & $D$ & 4$^{++}$ & 2 & 1$^{+}$ &  & $b_{1}\pi $ & $F$ & 4$^{++}$ & 1 &
1$^{+}$\\\cline{1-5}\cline{7-11}%
$\omega\rho $ & $F$ & 2$^{-+}$ & 1 & 0$^{+}$ &  & $\rho_{3}\pi $ & $S$ & 3$^{++}$ & 3 &
0$^{+}$\\\cline{1-5}\cline{7-11}%
\end{tabular}
\end{table}

In addition to these waves an isotropic non-interfering background wave was
included at each stage to account for the small waves that were omitted from
the fit, as well as the non-$\omega$ background. \ Lastly, a rank-1 fit with
the same wave set was compared with the rank-2 result. \ The wave intensities
were similar for the two fits, indicating that a rank-1 approximation was
adequate to describe the data.  The rank-1 results are discussed below.  
Mass distributions and angular distributions predicted from the
fitted amplitudes are in good agreement with the measured data. \ In this
letter we report the results for masses above the $\omega\rho^{-}$ threshold.  The 
data at lower masses are dominated by $a_2(1320)$ decay (see Figure~\ref{fig:1}(b)).
\ Further details of the analysis can be found in Ref.~\cite{Lu}.

\begin{figure}
\includegraphics[width=8.0cm]{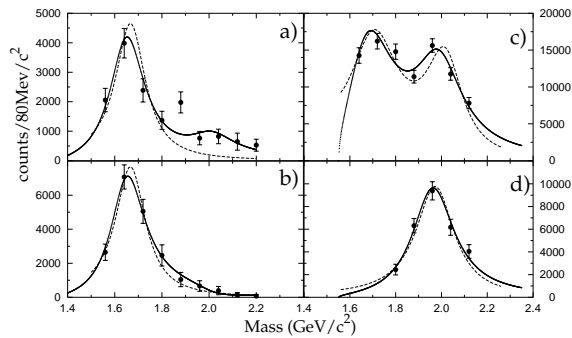}%
\caption{\label{fig:2}Wave intensity for (a) $1^{-+}(b_{1}\pi)^{S}_{1}1^{+}$, (b) $1^{-+}(b_{1}\pi)^S_10^{-}$, 
(c) $2^{++}(\omega\rho)^S_21^{+}$, and (d) $4^{++}(\omega\rho)^D_21^{+}$.
 The solid line is the Breit-Wigner result for two 1$^{-+}$ poles and the
dashed line is for one.}
\end{figure}

\begin{figure}
\includegraphics[width=9.0cm]{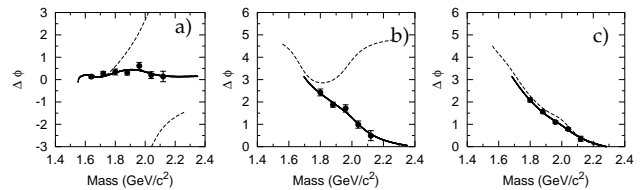}%
\caption{\label{fig:3}Phase difference for (a) $1^{-+}(b_{1}\pi)^{S}_{1}1^{+}-2^{++}(\omega\rho)^S_21^{+}$, 
(b) $1^{-+}(b_{1}\pi)^{S}_{1}1^{+}-4^{++}(\omega\rho)^D_21^{+}$
and (c) $2^{++}(\omega\rho)^S_21^{+}-4^{++}(\omega\rho)^D_21^{+}$. \ The solid
line is the Breit-Wigner result for two 1$^{-+}$ poles and the dashed line is
for one.}
\end{figure}

In the final phase of the analysis the PWA results for some of the largest waves were
fitted to linear combinations of relativistic Breit-Wigner poles.  
Mass-dependent resonance widths and Blatt-Weisskopf barrier factors were used~\cite{Kuhn}. 
Two separate fits were 
performed.  In the
first fit, shown in Figures~\ref{fig:2} and \ref{fig:3}, the 
intensities and phases
of the largest 1$^{-+}$, 2$^{++}$ and 4$^{++}$ waves were fitted, with common
resonance parameters in both natural and unnatural parity $1^{-+}$ waves.  Two 1$^{-+}$ 
poles were included in the fit.
\ The exotic $\pi_{1}(1600)$ was observed in the $b_{1}\pi$ channel, and $\omega\rho$ decay was 
measured for 
the previously identified $a_{2}(1700), a_{2}(2000)$, and $a_{4}(2040)$ states~\cite{pdg}.\ 
The resulting resonance parameters are given in Table~\ref{table:2}, with
statistical and systematic errors.  The quoted resonance widths are the fitted values uncorrected 
for resolution.
The systematic errors were determined by
repeating the resonance fits for PWA results with different wave sets and different mass binning, 
and using an alternative prescription for the mass dependent
width~\cite{Chungform}. \ Note that $a_{4}(2040)$ was observed with a smaller
width than expected, and at a lower mass than previously indicated~\cite{pdg}. 
The width of $\pi_{1}(1600)$ was measured with 
higher accuracy than previously and the value, 185$\pm25\pm
28$ MeV/$c^{2}$,  is
 smaller than that observed in $f_{1}\pi$~\cite{Kuhn}$\ $and $\eta^{\prime}\pi$~\cite{eta'}\ decay.

This fit also confirms the exotic $\pi_{1}(2000)$, a state previously discovered in
$f_{1}\pi$ decay~\cite{Kuhn}. \ In a fit without the $\pi_{1}(2000)$ pole,
$\chi^{2}$ increased from 30.7 (for 25 degrees of freedom) to 965 (for 31 degrees of freedom).  
That result is 
depicted as the dashed curve in Figures~\ref{fig:2} and \ref{fig:3}.  The mass of $\pi_{1}(2000)$, 
M $=2014\pm20\pm16$ MeV/$c^{2}$, is in good agreement with lattice gauge ~\cite{lattice} 
predictions for the 
lightest spin-exotic meson, as well as flux-tube model estimates for a hybrid meson~\cite{Isgur,Barnes95}  .

The
$\pi_{1}(1600)$ was observed in both natural and unnatural parity exchange,
with the largest strength in the unnatural parity wave. However $\pi
_{1}(2000)$ is excited primarily by natural parity exchange. \ Negligible
$\omega\rho^{-}$ resonance strength was observed for the exotic waves so they
were not included in the final fit. \ A large ratio of $b_{1}\pi$ to
$\omega\rho$ decay strength is expected for a hybrid meson~\cite{Isgur}.
\ Thus both $\pi_{1}(1600)$ and $\pi_{1}(2000)$ remain as hybrid meson
candidates as far as decay rates are concerned. \ However $b_{1}\pi$ decay is
predicted to dominate for hybrid $\pi_{1}$ decay, so one should expect primarily 
unnatural parity hybrid excitation with pion beams. \ Therefore the present
data favor a hybrid interpretation for $\pi_{1}(1600)$ based on the excitation
mechanism.  This result is at odds with the $f_{1}\pi$~\cite{Kuhn}
and $\eta^{\prime}\pi$~\cite{eta'} data since $\pi_{1}(1600)$ was observed only
in natural-parity exchange in those cases. \ Thus the data suggest that two
different $\pi_{1}$ states may have been observed at 1.6 GeV/$c^{2}$ (see
also Ref.~\cite{Sz}).
\begin{table}[tbp] \centering
\caption{Resonance parameters.  Here the subscript on the measured 
decay is the coupled intrinsic spin of the isobars.\label{table:2}}
\begin{tabular}
[c]{|l|l|l|l|}\hline
$\mathbf{resonance}$ & $\mathbf{decay}$ & $\mathbf{mass(MeV/c}^{2}\mathbf{)}$
& $\mathbf{width(MeV/c}^{2}\mathbf{)}$\\\hline
$a_{4}(2040)$ & $(\omega\rho)_{2}^{D}$ & 1985$\pm10\pm13$ & 231$\pm30\pm
46$\\\hline
$a_{2}(1700)$ & $(\omega\rho)_{2}^{S}$ & 1721$\pm13\pm44$ & 279$\pm49\pm
66$\\\hline
$a_{2}(2000)$ & $(\omega\rho)_{2}^{S}$ & 2003$\pm10\pm19$ & 249$\pm23\pm
32$\\\hline
$\pi_{1}(1600)$ & $(b_{1}\pi)_{1}^{S}$ & 1664$\pm8\pm10$ & 185$\pm25\pm
28$\\\hline
$\pi_{1}(2000)$ & $(b_{1}\pi)_{1}^{S}$ & 2014$\pm20\pm16$ & 230$\pm32\pm
73$\\\hline
$\pi_{2}(1670)$ & $(\omega\rho)_{1,2}^{P}$ & 1749$\pm10\pm100$ & 408$\pm
60\pm250$\\\hline
$\pi_{2}(1880)$ & $(\omega\rho)_{1,2}^{P}$ & 1876$\pm11\pm67$ & 146$\pm
17\pm62$\\\hline
$\pi_{2}(1970)$ & $(\omega\rho)_{1,2}^{P}$ & 1974$\pm14\pm83$ & 341$\pm
61\pm139$\\\hline
\end{tabular}
\end{table}

\begin{figure}
\includegraphics[width=9.0cm]{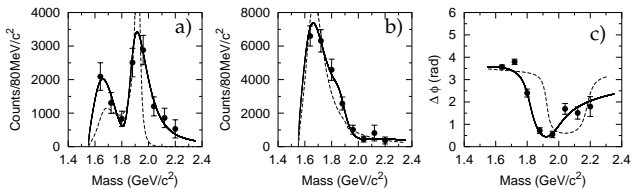}%
\caption{\label{fig:4}Wave intensity for (a) $2^{-+}(\omega\rho)^P_10^{+}$, and (b)
$2^{-+}(\omega\rho)^P_20^{+}$, and (c) phase difference for (a)$-$(b).  The solid
line is the Breit-Wigner result for three $2^{-+}$ poles and the dashed line is
for two.}
\end{figure}

The second fit was to the intensities and relative phase of 
the two largest 2$^{-+}$ waves.   Both
waves are natural-parity $\omega\rho P$ waves. Three resonance poles were used.  
The results of the fit are shown as the solid curve 
in Figure~\ref{fig:4}. \ This fit gave $\chi^{2}=9.0$ for 7 degrees of freedom.
Large $\omega\rho$ decay widths were observed for $\pi_{2}(1670)$ and for $\pi_{2}(1880)$, 
a state first observed by Anisovich, \textit{et al}.~\cite{Anisovich1}.  
Our value for the mass of $\pi_{2}(1880)$, M $=1876\pm11\pm67$
MeV/$c^{2}$, is in good agreement with the earlier measurement, M $=
1880\pm20$ MeV/$c^{2}$~\cite{Anisovich1}.  The isoscalar partner of this
state, $\eta_{2}(1870)$, is well known~\cite{pdg}. The presence of $\pi_{2}(1880)$ in the
spectrum prohibits the use of $\pi_{2}(1670)$ decay as a simple test of
hadronic decay models, as proposed by Page and Capstick~\cite{Pagemodel},
because there is significant mixing of $\pi_{2}(1670)$ with $\pi_{2}(1880)$.  

The $\pi_{2}$ fit included a third pole above the $\pi_{2}(1880)$, 
yielding $\pi_{2}(1970)$ with mass M $=1974\pm14\pm83$ MeV/$c^{2}$ and 
width $\Gamma=341\pm61\pm139$ MeV/$c^{2}$.   The $\pi_{2}$ data are poorly described 
in a fit without this 
resonance, as shown by the dashed curve in Figure~\ref{fig:4}.
High-lying $\pi_{2}$ strength was reported in several previous experiments.  
Measurements of $f_{1}\pi^{-}$ decay~\cite{Kuhn} 
revealed a resonance with mass M $=2003\pm88\pm148$ MeV/$c^{2}$ and 
width $\Gamma=306\pm132\pm121$ MeV/$c^{2}$, in good agreement with the present 
values.  A broad structure was also observed at 2.1 GeV/$c^{2}$ in three-pion decay~\cite{amelin2}. 
Those earlier measurements may include contributions from both $\pi_{2}(1880)$ and $\pi
_{2}(1970)$. Table~\ref{table:2} lists all of the resonance parameters from the present analysis.

One of the means by which unusual mesons can be identified is to measure a higher density 
of states than the quark model predicts.  In the quark model 
the $\pi_{2}(1670)$ is the ground-state $2^{-}$ 
configuration and
 the first radial excitation is expected at about 2.1 GeV/$c^{2}$~\cite{Godfrey}.  
This suggests a conventional 
meson interpretation for the $\pi_{2}(1970)$, leaving the $\pi_{2}(1880)$ as a hybrid meson
candidate.  The large $a_{2}\eta$ decay strength measured for $\pi_{2}(1880)$ also 
supports this assignment~\cite{Anisovich1,Kuhn}.  Thorough knowledge of the decay properties 
of $\pi_{2}(1880)$ and $\pi_{2}(1970)$ will aid in their identification~\cite{Isgur,Barnes}.  Further 
analysis of the present data, including the unnatural-parity $\pi_{2}$ waves listed in Table~\ref{table:1}, 
is now underway.

In summary, we observe strong excitation of the exotic $\pi_{1}(1600)$ in the
$(b_{1}\pi)^{-}$ decay channel, and confirm $\pi_{1}(2000)$.
Three $\pi_{2}$ states were measured between 1.5 and 2.2 GeV/$c^{2}$. In
addition to the well known $\pi_{2}(1670)$ we observe $\pi_{2}(1880)$ and $\pi_{2}(1970)$ decaying 
to $\omega\rho^{-}$. The higher state, $\pi_{2}(1970)$, is probably a radial excitation while 
the $\pi_{2}(1880)$ may have a large hybrid meson component.

This research was supported in part by the US Department of Energy, the US
National Science Foundation, and the Russian State Committee for Science and Technology.

\end{document}